# Mapping ultrafast timing jitter in dispersion-managed 89 GHz frequency microcombs via self-heterodyne linear interferometry

**Wenting Wang[1,2,†], Wenzheng Liu[1,*,†], Hao Liu[1,†], Tristan Melton[1], Alwaleed Aldhafeeri[1], Dong-IL Lee[1], Jinghui Yang[1], Abhinav Kumar Vinod[1], Jinkang Lim[1], Yoon-Soo Jang[1], Heng Zhou[3], Mingbin Yu[4,5], Patrick Guo-Qiang Lo[4,6], Dim-Lee Kwong[4], Peter DeVore[7], Jason Chou[7], Ninghua Zhu[8] and Chee Wei Wong[1,*]**

[1] Fang Lu Mesoscopic Optics and Quantum Electronics Laboratory, University of California, Los Angeles, CA 90095, United States of America

[2] Mesoscopic Optics and Advanced Instruments Laboratory, School of Optics and Photonics, Beijing Institute of Technology, Beijing, China

[3] Key Lab of Optical Fiber Sensing and Communication Networks, University of Electronic Science and Technology of China, Chengdu 611731, China

[4] Institute of Microelectronics, A*STAR, Singapore 117865, Singapore

[5] State Key Laboratory of Functional Materials for Informatics, Shanghai Institute of Microsystem and Information Technology, and Shanghai Industrial Technology Research Institute, Shanghai, China

[6] Advanced Micro Foundry, Singapore 117685, Singapore

[7] Lawrence Livermore National Laboratory, Livermore, CA 94550, United States of America

[8] Institute of Intelligent Photonics, Nan Kai University, Tianjin, China

[†] These authors contributed equally to this work.

**Abstract.** Laser frequency microcombs provide a series of equidistant, coherent frequency markers across a broad spectrum, enabling advancements in laser spectroscopy, dense optical communications, precision distance metrology, and astronomy. Here, we design and fabricate silicon nitride, dispersion-managed microresonators that effectively suppresses avoided mode crossings and achieves close-to-zero averaged dispersion. Both the stochastic noise and mode-locking dynamics of the resonator are numerically and experimentally investigated. First, we experimentally demonstrate thermally stabilized microcomb formation in the microresonator across different mode-locking states, showing negligible center frequency shifts and a broad frequency bandwidth. Next, we characterize the femtosecond timing jitter of the microcombs, supported by precise metrology of the timing phase and relative intensity noise. For the single-soliton state, we report a relative intensity noise (RIN) of -153.2 dB/Hz, close to the shot-noise limit, and a quantum-noise-limited timing jitter power spectral density (PSD) of 0.4 a.s$^2$/Hz at a 100 kHz offset frequency, measured using a self-heterodyne linear interferometer. Additionally, we achieve an integrated timing jitter of 1.7 ± 0.07 fs, measured from 10 kHz to 1 MHz. Measuring and understanding these fundamental noise parameters in high-clock-rate frequency microcombs is critical for advancing soliton physics and enabling new applications in precision metrology.

* To whom correspondence should be addressed. E-mail: wzliu@g.ucla.edu; cheewei.wong@ucla.edu

## 1 Introduction

Laser frequency combs have impacted science and technology fields with their equidistant frequency spacings, serving as unique coherent clockwork [1,2]. Recent emerging applications include, for example, clocks for space-borne networks [3,4], precise laser ranging metrology for autonomous platforms [5] and low phase noise radio frequency generation [6, 7], all aided by low timing jitter mode-locked frequency combs. The observations of dissipative soliton microcombs in single microresonators [8,9] or coupled-microresonators [10] with smooth spectral profiles and dispersive waves [11] offer opportunities to examine soliton comb dynamics in miniature platforms. There has been significant progress of soliton microcomb formation in different integrated microresonator platforms such as $Si_3N_4$ [12], AlN [13], $LiNbO_3$ [14], and AlGaAs [15], benefiting from either ultrahigh quality factors or large nonlinear coefficients. The recent demonstrations of electrically pumped turn-key soliton microcombs [16-18] and mode-locked microcombs [19] further reinforce the viability of the fully integrated frequency microcomb and pave the way for integrated functionalities such as terabit-per-second coherent transceivers [20-22], parallel coherent LiDAR [23], astrophysical spectrographs [24, 25], laser spectroscopy [26-28], distance ranging [29-31], low-noise microwave generation [32,33,36], and convolutional processing networks [34,35].

In soliton microcombs, the pump-resonance detuning noise [37] plays a critical role in the pump-to-repetition-rate noise transduction [38-40]. A low repetition rate phase noise (repetition rate timing jitter) regime exists at a detuning where the soliton center frequency shift from dispersive-wave emission is balanced by nonlinear effects. The phase noise can be improved by injection locking pump laser to resonant cavities [19], pumping the microresonators with a narrow linewidth laser [41], optimizing high-order dispersion of the microresonators [42], and thermal stabilization with an auxiliary laser [43]. Quantum motion of the microresonators has also been observed recently through timing jitter characterization in counter-propagating soliton pairs after suppressing common-mode technical noise [46]. With close-to-zero net group velocity dispersion, dispersion-managed soliton microcombs have been theoretically and experimentally investigated in active resonators featuring shorter pulse width as well as better timing stability [45-47]. Therefore, the precise characterization of timing jitter in various microcombs is highly demanded. Direct photon-detection can characterize timing jitter when repetition rates are detectable, but it has limited timing jitter power spectral density (PSD) noise floor of $1\times10^{-6}$ fs$^2$/Hz at 1 MHz offset frequency [12]. It is sensitive to intensity-noise-to-phase-noise (IM-PM) conversion [48]. Linear fiber interferometry [36,49] could provide a lower timing jitter PSD noise floor of $1\times10^{-9}$ fs$^2$/Hz which is free of the IM-PM conversion and shot noise limit.

Here we demonstrated a series of thermally intracavity-power-stabilized microcombs at different mode-locked states in 89 GHz dispersion-managed $Si_3N_4$ microresonators with negligible center frequency shift and broad frequency bandwidth. The demonstrated dispersion-managed (DM) microcombs not only expand the scope of soliton dynamics [8-11], but also enable low-jitter soliton trains. The DM-microcombs suppressed occurrences of avoided-mode crossings [50, 53, 55], compared to constant-dispersion microcombs, allowing lower timing jitter PSD. Subsequently, we determine the intensity and timing fluctuations of the soliton microcombs at the single-soliton, multiple soliton, and soliton crystal [51] states. Since microcomb oscillators have high repetition rates, low pulse energy, and high pulse background, we present a linear interferometry approach with tens of zeptosecond/Hz$^{1/2}$ timing jitter resolution to characterize its jitter. We note that the approach in [36, 49] is reference-free and independent of the repetition rate, expanding from prior silica microcomb [36] and fiber comb studies [49] to the 89 GHz pulse train timing jitter measurements of the silicon nitride DM-microcombs. The measurement of the fundamental timing jitter is based on: (1) time delay for the frequency discrimination and (2) optical carrier interference for optical phase discrimination. We observe a relative intensity noise (RIN) of -153.2 dB/Hz at 100 kHz offset, with a corresponding integrated RIN of 0.034 % from 100 Hz to 10 MHz for the single-soliton microcomb. For the single-soliton microcomb, the quantum-noise-limited timing jitter PSD is determined as 0.4 as$^2$/Hz for 100 kHz offset, with an integrated jitter of 1.7 ± 0.07 fs from 10 kHz to 1 MHz. The integrated timing jitter from 10 kHz to 44.5 GHz Nyquist is ≈ 32.3 fs.

## 2 Methods

### *2.1 Dispersion-managed microresonator fabrication*

The fabrication procedure of the microresonator starts with a 3 μm thick $SiO_2$ layer that is first deposited via plasma-enhanced chemical vapor deposition (PECVD) on a *p*-type 8” silicon wafer to serve as the under-cladding oxide. An 800 nm silicon nitride is subsequently deposited via low-pressure chemical vapor deposition (LPCVD) and the resulting nitride layer is patterned by optimized 248 nm deep-ultraviolet lithography and etched down to the buried oxide cladding via an optimized reactive ion etch. The nitride microresonators are then over-cladded with a 3 μm thick oxide layer, deposited initially with LPCVD for 0.5 μm and then with PECVD for 2.5 μm.

*2.2 Dispersion-managed microcomb formation numerical simulation and stochastic noise calculation*

After taking anomalous GVD and AMX into consideration, we have implemented the Ikeda method to obtain the roundtrip-varying non-mean-field microcomb dynamics and noise character written as:

$$\begin{cases} E_{m+1}(0,t) = \sqrt{T_c}E_{in} + \sqrt{1-T_c}e^{-i(\omega_0-\omega_p)t_R}E_m(L_{cav},t) \\ \dfrac{\partial \tilde{E}_m(z,\omega)}{\partial z} = -\left[\alpha - i\dfrac{\beta_2(z)}{2}\omega^2\right]\tilde{E}_m(z,\omega) + i\gamma(z)\tilde{E}_m(z,\omega) \star \{\mathcal{F}[R(t')] \times \mathcal{F}[|E_m(z,t)|^2]\} \end{cases}$$

where $t_R$ is the round-trip time, $z$ is the propagation distance within each cavity round trip, $t$ is the fast time, $E_m(z,t)$ is the intracavity electric field of the m$^{th}$ roundtrip, the fourier transform correspondence on the fast time is $\tilde{E}_m(z,\omega)$, $E_{in}$ is the external pump field, $\alpha$ is the propagation loss, $T_c$ is the coupling strength, $L_{cav}$ is the cavity length, $\delta = (\omega_0 - \omega_p)t_R$ is the pump-resonance detuning phase, and $\beta_2(z)$ is the second order dispersion coefficient varying over the roundtrip position. Here, $\beta_2(z)$ is chosen from Figure 1b ensuring the averaged $\beta_2$ is -4.3 fs$^2$/mm, based on our experimental characterization. $\gamma(z)$ is the Kerr nonlinear coefficient varying over the roundtrip position. $R(t)$ is the nonlinear response function, including the Raman response function. To incorporate the avoided-mode-crossing-induced frequency shift, an additional frequency shift $\Delta_n$ is introduced to the $n_{th}$ mode, so that the mode frequency becomes: $\omega_n = \omega_0 + D_1 n + \frac{D_2 n^2}{2} + \Delta_n$. $\Delta_n$ is determined by the empirical two-parameters model $\frac{\Delta_n}{2\pi} = \frac{-a/2}{n-b-0.5}$, where $a$ is the maximum mode frequency shift and $n$ and $b$ are the mode number and mode number for the maximum mode frequency shift, respectively. Thermal effect is not considered in the simulation because of the dual-driven thermal balance and the TEC module. From the cavity mode dispersion characterization, the cavity free-spectral range $D_1/2\pi$ = 88.52 GHz. Furthermore, we estimate the maximum frequency shift to be $\frac{\Delta_n}{2\pi} = 130$ MHz at 1581.5 nm determined by the spectral peak in the soliton microcomb optical spectrum. The estimated value is supported by comparing the simulated comb spectrum with the experimental result, which are in good qualitative agreement. 2,000 modes centered at the pump are incorporated in the model. The simulation starts with vacuum noise and runs for $1 \times 10^5$ roundtrips until the solution reaches steady state.

To characterize the performance of the timing jitter, we introduced the thermal noise $\epsilon_\Theta$, shot noise $\epsilon_{\mathrm{s}}$, pumping laser's intensity noise $\epsilon_{\mathrm{in}}$ and frequency noise $\epsilon_\omega$ into our modeling. The dynamics can be represented as

$$\begin{cases} E_{m+1}(0,t) = \sqrt{T_c}(E_{in}+\epsilon_{\mathrm{in}}) + \sqrt{1-T_c}\, e^{-i(\omega_0-\omega_p-\epsilon_\omega)t_R} E_m(L_{cav},t) + \epsilon_{\mathrm{s}} \\ \dfrac{\partial \tilde{E}_m(z,\omega)}{\partial z} = -\left[\alpha - i\dfrac{\beta_2(z)}{2}\omega^2 - i\epsilon_\Theta\right]\tilde{E}_m(z,\omega) + i\gamma(z)\tilde{E}_m(z,\omega) \star \{\mathcal{F}[R(t')] \times \mathcal{F}[|E_m(z,t)|^2]\} \end{cases}$$

All these noise sources are assumed proportional to the normal distribution. The standard derivations of $\epsilon_{\mathrm{in}}$ and $\epsilon_\omega$ are determined by the laser's RIN PSD and frequency noise PSD, respectively [52, 61]. To extend simulated Fourier frequency range to kHz frequencies, we examine this for a total of $2^{24}$ roundtrips.

*2.3 Relative intensity noise and self-heterodyne linear interferometry*

The filtered microcomb after removing the pump laser is measured by a photodetector (Thorlabs PDA10CF) with optical power of 210 μW. A multimeter and an oscilloscope monitor the DC voltage ($V_0$). A signal source analyzer (Keysight E5052B) records the voltage fluctuation PSD $S_{\Delta I}(f)$. A home-built diffractive grating pair is used to select the comb lines for the timing jitter PSD measurement. We first optimize the noise floor of the SHLI by improving the signal-to-noise ratio of the detected RF signal at 100 MHz. Secondly, we minimize the relative delay time between the two optical comb lines ($\nu_{\mathrm{n}} = 190.11$ THz, $\nu_{\mathrm{m}} = 192.55$ THz) and power difference of the two arms of SHLI, to enable the suppression of the common-mode noise. Thirdly, we optimize the delay time for the soliton microcomb based on two criteria: the first is to maximize the timing jitter measurement sensitivity and the second is to expand the measured Fourier offset frequency range. The fiber delay length was optimized to be 49 m. The detected RF power at the output of the two photodetectors are 14 and 19 dBm.

**3 Results**

*3.1 Chip-scale low timing jitter dispersion-managed silicon nitride microresonators*

Figure 1a shows the schematic illustration of the DM soliton microcomb formation in the tapered microresonator which includes scanning electron microscope images of the stoichiometric silicon nitride microresonator with 261 μm outer radius and 800 nm thickness. The nitride waveguide width is continuously changed from 1 to 4 μm to finely tune group velocity dispersion and filter high-order transverse modes in the single-mode microresonator [51]. Group velocity dispersion (GVD) varies along the microresonator from -55 to 58 fs²/mm simulated with the finite-element analysis method as shown in Figure 1b after considering both geometric and material dispersion. Figure 1c shows the simulated GVD for various waveguide widths along with the cavity-path averaged GVD calculated via the relation $\beta_2 = \int \beta_2(z) dL / L_{cavity}$ where $\beta_2(z)$ is the GVD for the microresonator waveguide, and $L_{cavity}$ is the cavity circumference.

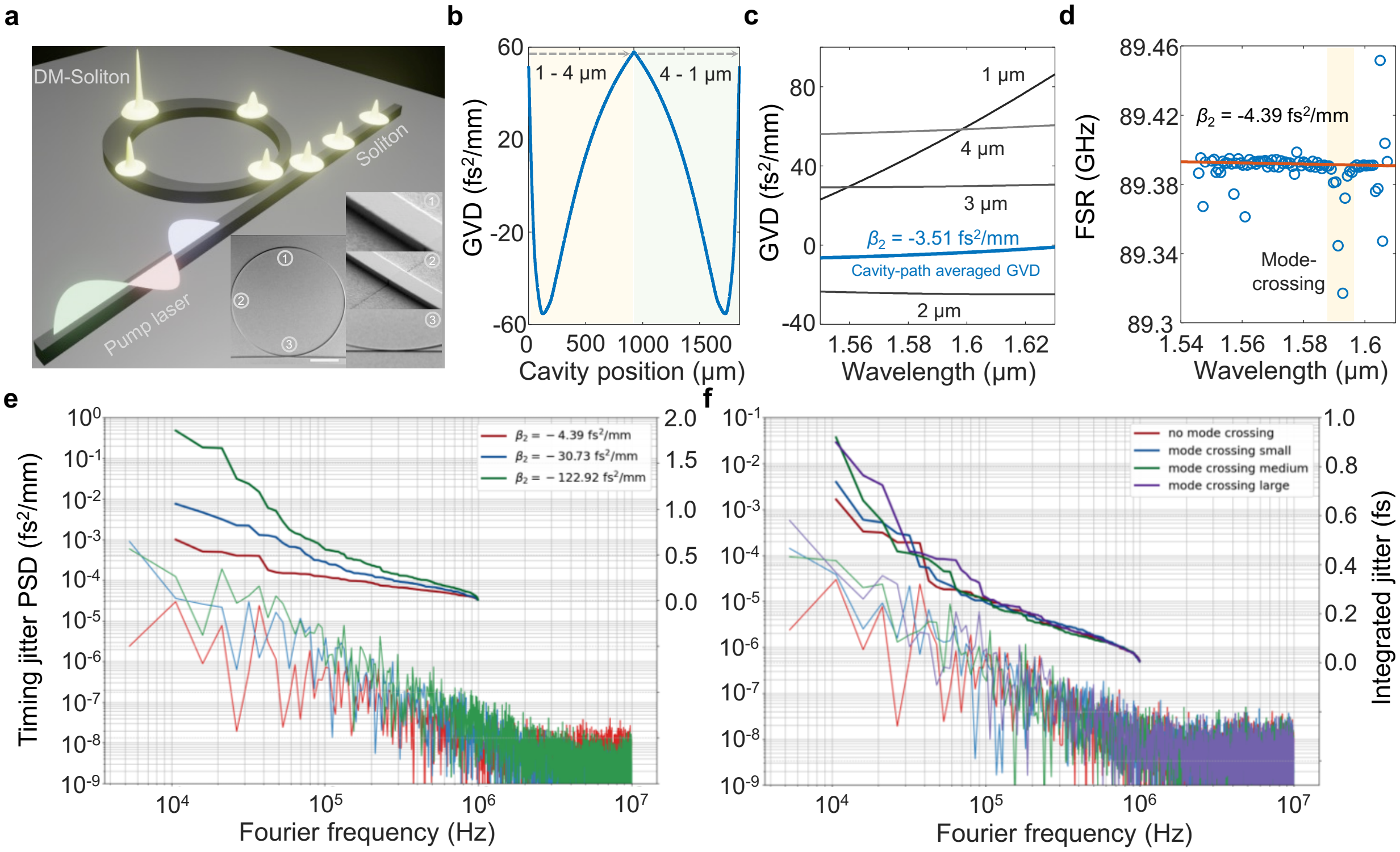

**Figure 1 | Chip-scale low timing jitter dispersion-managed silicon nitride microresonator. a,** Schematic illustration of dispersion-managed (DM) soliton microcomb generation. Inset: scanning electron microscope images of the microresonator, zoomed waveguides, and coupling gap. Scale bar: 130 μm. **b,** Simulated group velocity dispersion (GVD) along the microresonator. **c,** Simulated GVD of waveguides with different widths at a fixed waveguide height and the cavity-path averaged GVD. **d,** Measured GVD via swept-wavelength interferometry showing $\beta_2$ = -4.39 fs²/mm. **e,** Modeled timing jitter power spectral density and integrated jitter versus Fourier frequency for a constant-dispersion uniform-width ring microcomb, with $\beta_2$ = -4.39 fs²/mm, -30.73 fs²/mm, and -122.92 fs²/mm respectively, no Raman and thermal effects assumed. **f,** Modeled timing jitter power spectral density

and integrated jitter versus Fourier frequency for a constant-dispersion uniform-width ring microcomb with different mode crossing levels.

The fundamental transverse electric mode features a small anomalous path averaged simulated GVD of -3.51 $fs^2$/mm at a pump wavelength of 1602 nm. The cavity GVD is experimentally characterized by swept-wavelength interferometry with only two avoided mode-crossings (AMX) across the entire wavelength range for near-single-mode operation. The measured free spectral range and GVD are ≈ 89 GHz and -4.39 $fs^2$/mm, respectively as shown in Figure 1d. These values deviate from the calculated dispersion due to factors such as fabrication imperfections, material inhomogeneity, and limitations in the accuracy of our measurement setup. The measured loaded and intrinsic quality factors are $1.8 \times 10^6$ and $3.4 \times 10^6$, respectively. Subsequently, we numerically examined the timing jitter PSD depending on the different intrinsic cavity GVD assumed no AMX, thermal, higher-order dispersion, and Raman effects shown in Figure 1e. As the GVD decreases, the jitter PSD is decreased. AMX will cause higher timing jitter PSD and higher integrated jitter accordingly, which will be proved in the numerical simulations as shown in Figure 1f. These results show the low jitter pulse train can be realized by close-to-zero net dispersion and suppression of AMX. Our dispersion-managed adiabatic rings are able to suppress the occurrence of these avoided-mode crossings and obtain small dispersion and thus reduce the timing jitter.

*3.2 Thermally stabilized dispersion-managed microcomb formation*

DM-microcombs are subsequently generated in the microresonator. Figure 2a illustrates the optical spectrum of the single-soliton DM-microcomb overlapped with the numerically modeled spectral profile. Modest spectral dips resulting from two hybridized inter-polarization mode couplings at 1592.64 nm and 1659.72 nm are observed. The 1563.64 nm peak is the auxiliary pump laser. The effect of the auxiliary pump diminishes once the soliton state is reached. To maintain stability, thermal noise needs to be actively controlled through feedback mechanisms. To illustrate the temporal performance of the microcomb, we measured the intensity autocorrelation (AC) trace with a non-collinear second-harmonic autocorrelator after pump suppression with a bandpass filter. Figure 2b shows the measured pulse width of the single-soliton at ≈ 305 fs for the filtered optical spectrum, along with the ≈ 11.2 ps pulse train. The modeled pulse width is included in the inset of Figure 2d. We also observed double-soliton and soliton crystal states in the microresonator. The corresponding measured optical spectra of the double-soliton and one defect soliton crystal are illustrated in Figure 2e and 2g overlapped by the modeled spectral profiles, respectively. The soliton crystal optical spectrum indicates destructive interference between a single-soliton microcomb and a 12-FSR perfect soliton crystal microcomb. The spatiotemporal modeled intracavity waveforms are depicted in Figure 2f and 2h where the soliton defect in the time domain is presented. The demonstrated microcombs offer broader optical spectra consistent with the simulated results over multiple soliton types with near-single-mode operation with only two avoided mode crossings and negligible center frequency shift [54] compared to prior studies [8-19].

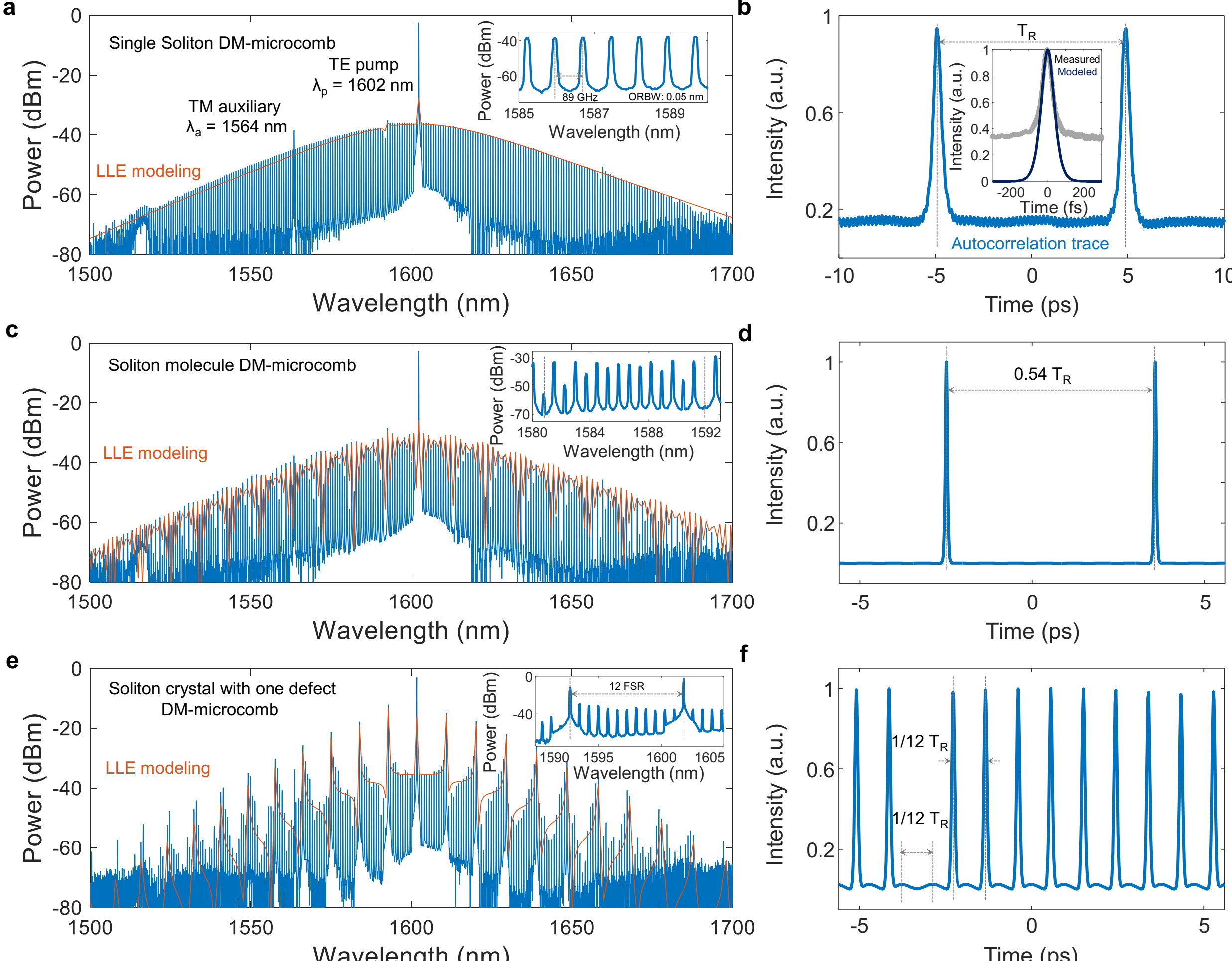

**Figure 2 | Soliton microcomb formation in tapered dispersion-managed microresonators. a, c,** and **e,** Measured optical spectra of the single-soliton, double-soliton and one defect soliton crystal DM-microcombs overlapped with the numerical model showing negligible center frequency shifts (detailed in Supplementary Information II). Insets are zoomed optical spectra. **b,** Measured intensity autocorrelation trace of the single-soliton DM-microcomb. Inset is the measured and modeled pulse width. **d,** Modeled intracavity waveform of the double-soliton with a temporal separation of 0.54 × $T_R$. **f,** Modeled intracavity waveform of the one defect soliton crystal showing the modulated background ("potential well") with a period of 1/12 × $T_R$.

*3.3 Relative intensity noise measurement of the dispersion-managed microcombs*

To obtain the DM-microcombs reliably and deterministically, Figure 3a illustrates the implemented TE-TM dual-driven pump approach (more details in Supplementary Information Section I). The forward-propagating pump laser is amplified and polarized into the transverse-electric (TE) polarization while the backward-propagating transverse-magnetic (TM) polarized auxiliary laser thermally stabilizes the microresonator intracavity total power. Through this approach, we generate the microcombs in a planar tapered dispersion-managed $Si_3N_4$ microresonator at the effective red-detuned region of resonance $v_{\mu}$ based on dynamic photothermal stabilization. The orthogonally polarized auxiliary laser is blue-detuned from the resonant mode $v_{\mu+53}$ in the reverse direction to mitigate thermal transients during the microcomb transition from a high-noise chaotic state to a low-noise mode-locking state and stabilize the pump-resonance detuning thermally. The TM-polarized auxiliary laser experiences normal GVD and avoids the initiation of parametric oscillation while also stabilizing intracavity power via optimization of power and phase detuning [43]. To quantify the soliton microcomb noise performance, we conduct intensity noise at different soliton microcomb states prior to the respective timing jitter measurements. Coherence of the soliton frequency microcomb is examined via the RF intensity noise spectra over a microwave frequency span that is a few times of the cold cavity resonance linewidth.

To supplement the RF intensity noise measurements of the soliton microcombs, measurement of the RIN is performed next. A signal source analyzer records the intensity fluctuation PSD of the soliton microcombs after suppressing the pump laser (detailed in Methods). The filtered optical spectrum is shown as in Figure 3b. The measured RIN PSD [$S_{RIN}(f)$] of the 89 GHz mode-locked microcombs, calculated by normalizing the measured intensity fluctuation PSD [$S_{\Delta V}(f)$ in units of $V^2$/Hz] by the average detected intensity $|V_0|^2$, is shown in Figure 3c to 3e corresponding to the single-soliton, double-soliton, and one defect soliton-crystal microcombs. The black curves in Figure 3c to 3e are the RIN PSD of the spatiotemporal chaotic state [52] [corresponding optical spectra is illustrated in Supplementary Information Section I, Figure S1b$_2$] and the pump laser, indicating the upper and lower bounds of the soliton microcomb RIN PSD. The soliton microcomb RIN PSD drops with a 30-dB/decade slope ($1/f^3$) over the first offset frequency decade while the continuous wave (CW) pump laser RIN PSD falls with a 20-dB/decade slope ($1/f^2$). The pump laser RIN PSD is measured at a non-resonant wavelength after the microresonator. The discrepancy between the two slopes is attributed to environmental noise sources such as free-space-to-chip coupling fluctuations. For the single-soliton state, the measured RIN is -153.2 dB/Hz at 100 kHz offset with a corresponding integrated RIN of 0.034% when integrated from 100 Hz to 10 MHz with the relation $RIN_{in} = \int_{f1}^{f2} S_{RIN}(f)\, df$ where $f_1$ and $f_2$ are the lower and upper offset frequency bounds. The measured RIN of the double-soliton state and the soliton crystal state are -149.8 dB/Hz and -148.6 dB/Hz respectively for a 100 kHz offset. The corresponding integrated RIN are 0.036% and 0.023% over the same integrated frequency range. The inset of Figure 3c shows the electrical noise suppression to facilitate the observation of the dynamical intensity noise of the microcombs by optimizing the incident power of the photodetector from -7.16 to -4.33 dBm. The insets of Figure 3d and 3e show the noise degradation of the double-soliton and soliton crystal microcombs which results from the conversion of phase fluctuations to intensity fluctuations in the intracavity spectral interference process.

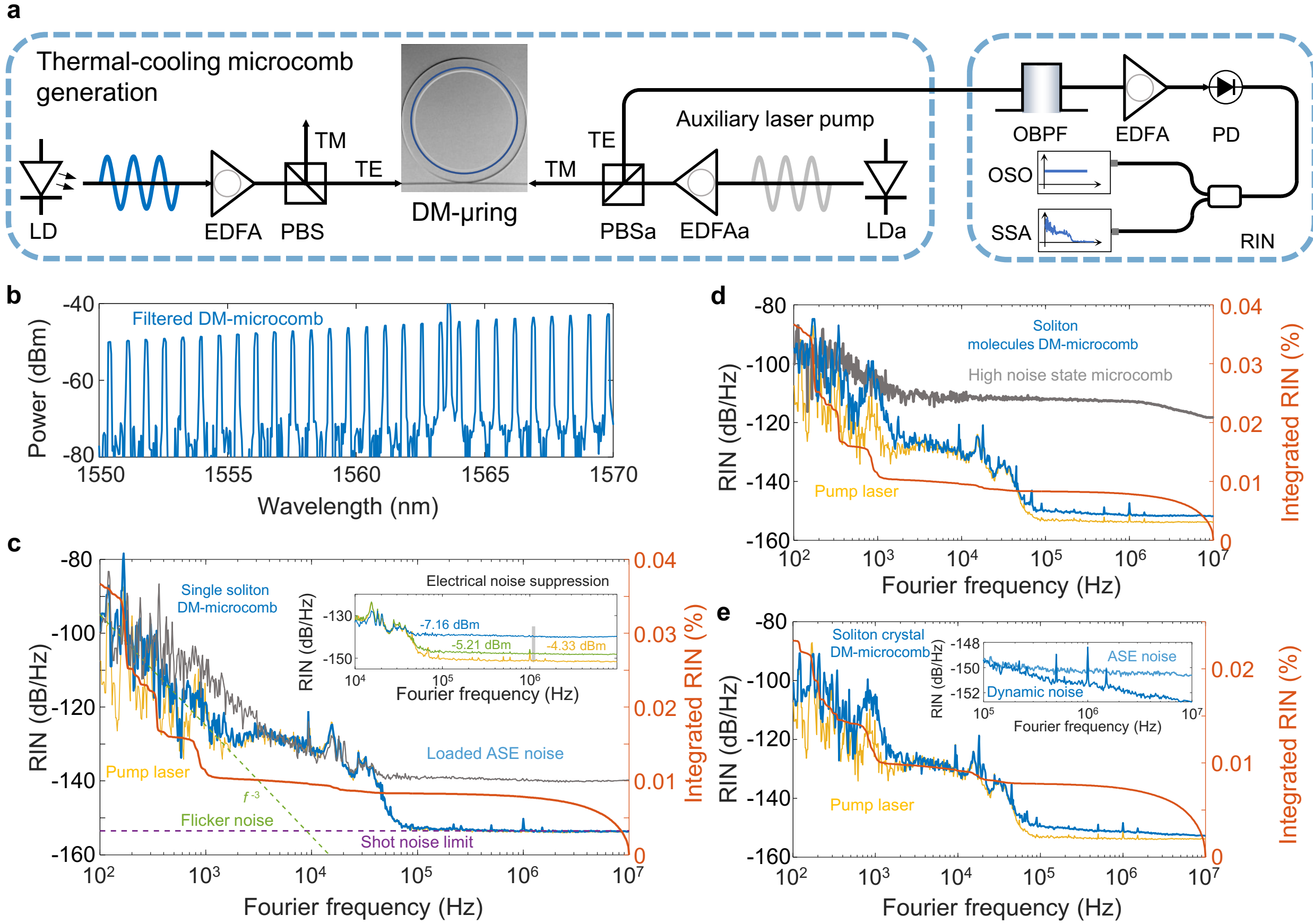

**Figure 3 | Relative intensity noise (RIN) measurements of the dispersion-managed microcombs. a,** Experimental setup of the TE-TM dual-driven approach for the generation of thermally stabilized soliton microcomb and the relative intensity noise measurement. LD: laser diode; EDFA: erbium-doped fiber amplifier; PBS:

polarization beam splitter; TE: transverse-electric; TM: transverse-magnetic, OBPF: optical bandpass filter; PD: photodiode; OSO: oscilloscope; SSA: signal source analyzer. **b,** Filtered optical spectrum of the single-soliton DM microcomb. **c, d,** and **e,** Relative intensity noise power spectral density (PSD) and the corresponding integrated RIN of the microcombs at the different dynamical states along with the lower bound set by the pump laser. The RIN PSD of the chaotic DM-microcomb and the RIN PSD after loading broadband ASE noise are also illustrated. Inset of **c**: Electrical noise optimization by adjusting the incident optical power of the PD to explore the dynamic soliton intensity fluctuations at the different states. Inset of **d**: Noise degradation of the double-soliton DM-microcomb showing additional white high-frequency noise. Inset of **e**: Noise degradation of the soliton crystal DM-microcomb showing dynamic high-frequency noise.

### *3.4 Self-heterodyne linear interferometry (SHLI) for soliton microcomb femtosecond jitter metrology*

We next examine the timing jitter via a self-heterodyne linear interferometer (SHLI). Figure 4a illustrates the implemented SHLI architecture for precision timing jitter metrology. The interferometer consists of a reference arm and a time-delayed arm in which a fiber-coupled acousto-optic modulator (AOM) driven by a RF signal $f_m$ allows heterodyne detection of phase fluctuation power spectral density (PSD, $S_\varphi$ [dB rad$^2$/Hz]). A diffractive grating-based narrowband filter pair selects two microcomb lines at frequencies $\nu_n = n \times f_R + f_{CEO}$ and $\nu_m = m \times f_R + f_{CEO}$ as illustrated in the inset of Figure 4a separated by a $(m\text{-}n) \times f_R$ frequency difference, where $f_R$ is the repetition rate and $f_{CEO}$ is the carrier envelope offset frequency of the microcombs. Frequency noise $\Delta\nu(f)$ of the selected lines is subsequently discriminated with delay time $\tau$ by the relation $\Delta\phi(f) = 2\pi\Delta\nu(f)\tau$. The optical phase fluctuations $\Delta\phi(f)$ are converted into optical intensity fluctuations by linear optical interferometry. At the fiber interferometer output, the two optical lines are demodulated by two photodetectors. The residual phase noise PSD originates from the frequency difference $(m - n) \times f_R$ of the two selected comb lines, proportional to delay time $\tau$. Then, a double-balanced mixer is used to extract the timing jitter PSD and eliminate the common-mode noise induced by the carrier envelope offset signal and the driven microwave frequency signal ($f_m$). The frequency fluctuations of the two selected optical comb lines are converted into voltage fluctuations with the transfer function $\Delta V(f) \propto K_\varphi \frac{|1-e^{-i2\pi f\tau}|}{|i\times f|}(m-n)\Delta f_R(f)$ where $K_\varphi$ is the peak voltage at the double-balanced mixer output. The transfer function shows that the measured voltage fluctuation is proportional to $\left|1 - e^{-i2\pi f\tau}\right|/|i \times f|$, which implies there will be null points at the offset frequency $f = 1/\tau$ and its harmonics, providing the upper Fourier frequency limit of the timing jitter measurement.

The measured voltage fluctuation PSD is subsequently converted into frequency noise PSD and further into timing jitter PSD $S_{\Delta T_R}(f)$ with the relation $S_{\Delta T_R}(f) = \left(\frac{1}{2\pi f_R}\right)^2 \frac{1}{(m-n)^2} \frac{1}{f^2} \left(\frac{|i \times f|}{K_\varphi |1-e^{-i2\pi f\tau}|}\right)^2 S_{\Delta V}(f)$. The detected voltage fluctuation at the mixer output is separated into two parts. The first part synchronizes the fiber interferometer to the frequency microcombs, avoiding free walk via a piezoelectric-transduced fiber stretcher (FS) through a loop filter with 1 kHz bandwidth. The second part is recorded by a signal source analyzer which gives the timing fluctuation PSD and the frequency fluctuation PSD of the soliton microcomb repetition rate. To precisely remove the common-mode noise resulting from dispersion and increase the interferometer signal-to-noise ratio, we utilize a delay control unit (DCU) which contains a motorized fiber delay line (MDL) and a pair of wavelength-division multiplexed (WDM) couplers. This timing-stabilized and dispersion-compensated fiber interferometer can be considered as a true time delay, which is an optical counterpart of the delay-line frequency discriminator in microwave metrology [56].

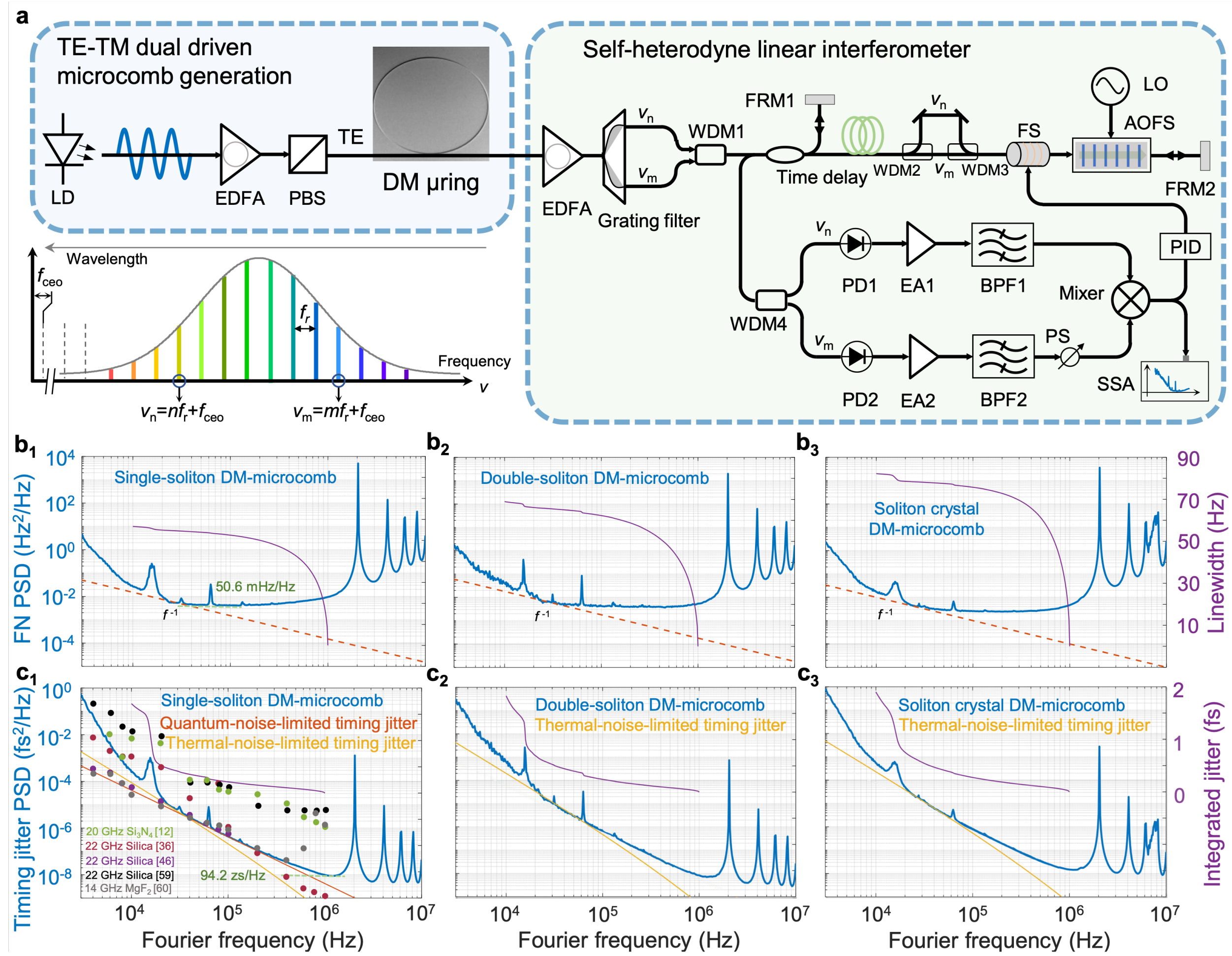

**Figure 4 | Measured repetition rate frequency noise PSD and timing jitter PSD and the corresponding integrated RF linewidth and timing jitter of dispersion-managed soliton microcombs. a,** Experimental setup of the self-heterodyne linear interferometer (HLI). WDM: wavelength division multiplexer; FS: fiber stretcher; EA: electronic amplifier; BPF: bandpass filter, PS: phase shifter; PID: proportional-integral-differential controller. Inset is the schematic illustration of the self-heterodyne linear interferometry. **b**$_1$**, b**$_2$**,** and **b**$_3$**,** Measured frequency noise PSD at different soliton states with a 49 m stabilized fiber link. The orange dashed lines with 10 dB/decade slopes indicate repetition rate frequency free-walk induced by microresonator intracavity power fluctuations. The corresponding repetition rate tone linewidth integrated from 1 MHz to 10 kHz is denoted with purple curves. The repetition rate carrier frequency is 89 GHz. **c**$_1$**, c**$_2$**,** and **c**$_3$**,** Timing jitter PSD measurement of the soliton microcombs at different dynamical states with the calculated thermal noise and quantum noise limits. The timing jitter theoretical models from Ref. 44 and Ref. 58 are quantum-noise and thermal-noise limits of the soliton microcombs denoted respectively with solid orange and yellow lines. The corresponding integrated timing jitter is included. The comparison between the measured timing jitter PSD and prior works [12,36,46,59,60] are included as well.

We convert the measured voltage fluctuation PSD on the baseband into the repetition rate frequency noise $S_{\Delta f_R}$ PSD to examine the frequency noise behavior as shown in Figure 4b$_1$, 4b$_2$, and $4b_3$ of the soliton microcomb at the single-soliton, double-soliton, and soliton crystal states, respectively. The measured repetition rate frequency noise PSD are 2556 mHz$^2$/Hz, 4151 mHz$^2$/Hz and, 4168 mHz$^2$/Hz respectively at 100 kHz offset with a noise frequency resolution of 64 mHz/Hz$^{1/2}$ at the single-soliton microcomb. We observe that the repetition rate frequency noise features a 20-dB/decade slope below ≈ 20 kHz offset, indicating the repetition rate random walk frequency noise. Based on the noise power-law, soliton microcombs have a repetition rate flicker frequency walk from ≈ 20 kHz to 40 kHz offset, a white frequency noise from ≈ 40 kHz to 200 kHz offset, and a flicker and white phase noise from ≈ 200 kHz to 1 MHz. The resulting integrated linewidths of the free-running repetition rate tone are shown in Figure 4b for the three microcomb dynamical states. We also note the repetition rate close-to-carrier phase noise can be stabilized to a low-noise microwave oscillator [24].

Figure $4c_1$, $4c_2$, and $4c_3$ show the measured timing jitter PSD for the different soliton dynamical states. For the single-soliton comb, the measured quantum-noise-limited timing jitter PSD is 0.4 as$^2$/Hz at 100 kHz offset. The corresponding integrated timing jitter is 1.7 ± 0.07 fs when integrated from 10 kHz to 1 MHz as shown in the Figure $4c_1$ which is close to the timing jitter in silica microresonator frequency microcombs measured with similar technology [36]. The integrated timing jitter from 10 kHz to 44.5 GHz Nyquist is ≈ 32.3 fs. The achieved femtosecond-level jitter is enabled by close-to-zero intracavity dispersion to minimize group delay fluctuations, suppressed Kerr nonlinearities within the tapered waveguide [45], and the thermally stabilized dual-driven approach. Our dispersion-managed microresonator stretches the soliton pulse within the cavity, reducing the accumulated nonlinear phase shift during pulse propagation. The quantum-noise-limited timing jitter PSD of the two (double-soliton and soliton crystal) states at 100 kHz offset are at 0.66 as$^2$/Hz and 0.82 as$^2$/Hz, respectively. This corresponds to an integrated jitter of 1.9 ± 0.06 fs and 1.8 ± 0.09 fs. The dynamical noise is observed in the timing jitter PSD of the soliton crystal microcomb at the offset frequency around 8 MHz. Compared to direct photon detection for timing jitter PSD measurements [12], the SHLI method can effectively avoid the intensity-noise-to-phase-noise (IM-PM) conversion and shot-noise limit. The measured timing jitter PSD compares to prior works [12,36,46,58,59] as shown in Figure $4c_1$.

For each of the microcomb soliton states, we observe that the timing jitter PSD drops with a 40-dB/decade slope within 3 to 20 kHz as shown in Figure 4c. Deviation of timing jitter PSD over low Fourier frequency is associated with intracavity power fluctuation leading to a $1/f^4$ slope with the relation $S_{\Delta T_R}^{T_R}(f) = \left(\frac{1}{2\pi f T_R}\right)^2 S_{T_R}(f)$ where intracavity power-induced round-trip fluctuations $S_{T_R} \propto f^{-2}$.

Figure 4c also plots the theoretical cavity thermal bounds on the timing jitter PSD with the yellow solid line, arising from thermo-refractive variation and bounding the measured timing jitter PSD from 20 kHz to 40 kHz with a 25-dB/decade slope ($1/f^{2.5}$). In the DM microresonator, the thermal-noise limited timing jitter PSD, originating from the thermodynamic fluctuations $\langle \delta T^2 \rangle = k_B T^2 / C V \rho$ where $V$ is the optical mode volume, $k_B$ is the Boltzmann constant, $T$ is the chip temperature, $\rho$ is density, and $C$ is specific heat capacity, is described with the model [57]

$$S_{\Delta T_R}(f) = \frac{1}{(2\pi f_R)^2} \left(\frac{v_c}{f_R} \frac{1}{n_0} \frac{dn}{dT}\right)^2 \frac{1}{f^2} \frac{k_B T^2}{\sqrt{2\pi^4 \kappa \rho C f}} \frac{1}{R\sqrt{d_r^2 - d_z^2}} \frac{1}{\left[1 + (2\pi f \tau_d)^{3/4}\right]^2} \tag{1}$$

where $dn/dT$ is the thermorefractive coefficient, $\kappa$ is thermal conductivity, $R$ is the microresonator ring radius, $d_{r(z)} = \int d_{r(z)} dL / L_{cavity}$ is the half-width of fundamental mode along the tapered DM microresonator, and $\tau_d = \frac{\pi^{1/3}}{4^{1/3}} \frac{\rho C}{\kappa} d_r^2$. From 40 kHz to 600 kHz, the measured PSD falls with a quantum-noise limited 20-dB/decade slope ($1/f^2$). The theoretical quantum-noise timing jitter limit without shot-noise shows in Figure 4c with orange solid lines following the model [44]

$$S_{\Delta T_R}(f) = \frac{1}{4\pi\sqrt{2} f_R^2} \sqrt{\frac{\gamma}{\Delta_0 D}} \frac{g}{\gamma^2} \left[\frac{1}{96} \frac{\gamma D}{\Delta_0} \frac{\gamma^2}{f^2} + \frac{1}{24} \left(1 + \frac{\pi^2 f^2}{\gamma^2}\right)^{-1} \frac{\gamma^2}{f^2} \frac{\Delta_0 D}{\gamma}\right] \tag{2}$$

where $\gamma$ is the half linewidth half height of the cavity resonance, $f_R$ is the repetition rate of the microcombs, $g = \frac{n_2}{n_0} \frac{\hbar \omega_c^2 c}{V n_0}$ is the nonlinear gain coefficient, $n_0$($n_2$) is the refractive index (nonlinear index) of the nitride resonator, $\omega_c = 2\pi v_c$, $v_c$ is the center frequency of the microcombs, $c$ is the light speed in vacuum, $D = -\frac{\beta_2 \omega_R^2 c}{\gamma n_0}$ is the normalized dispersion and $\Delta_0 = \omega_0 - \omega_P$ is the resonance-pump detuning. Above 600 kHz, the measured PSD is limited to 8905 zs$^2$/Hz by the SHLI spectral resolution.

Based on soliton theory, the quantum-noise-limited timing jitter PSD model, especially in the high offset frequency more than 10 kHz, can analytically predict noise behaviors for different mode-locked states via the relation $S_{\Delta T_R}(f) \approx 0.5294 \frac{\xi}{(2\pi f)^2} \frac{h\nu}{E_p} \frac{\alpha_{tot}}{T_R} \tau_p^2$ [58], where $E_p \approx$ $\frac{4\pi h\theta v_c}{D_1\gamma_c}\sqrt{2D_2\Delta_0}$ is the intracavity pulse energy, $D_1/2/\pi$ is the cavity FSR, $D_2$ is related to cavity GVD, $\theta$ is the transmission of the microresonator, $\gamma_c$ is the cubic nonlinearity parameter, $\tau_p \approx \frac{1}{D_1}\sqrt{\frac{D_2}{2\Delta_0}}$ the intracavity pulse duration [37], ξ and $\alpha_{tot}$ are the spontaneous emission factor and cavity loss. For the different soliton states, the quantum-noise-limited timing jitter PSD is inversely proportional to the resonance-pump detuning and proportional to square root of the cavity dispersion. The soliton microcomb center frequency fluctuation PSD $S_{\Delta v_c}(f)$ can also be converted into the timing jitter with the relation $S_{\Delta T_R}(f) \approx \left(\frac{D_2}{fT_R}\right)^2 S_{\Delta v_c}(f)$ where $\Delta v_c$ is the center frequency fluctuations induced by avoided-mode-crossings [39,40], odd-order dispersion [42], or Raman effects [54]. In addition, the intracavity intensity fluctuations will introduce the extra timing jitter PSD with the relation of $S_{\Delta T_R}(f) = C \times (\eta P_{in})^2 \left(\frac{1}{f}\right)^2 S_{RIN}(f)$ where $\eta = df_R/dP_{in}$ is the transduction factor, $P_{in}$ is the microresonator intracavity power, $C$ is a constant [36, 43].

We note our noise measurements of the frequency microcombs below the offset frequency of 20 kHz are still higher than the microresonator theoretical thermodynamical limits. This is attributed to the strong free-running intracavity power fluctuations and pump-resonance detuning noise. Further active stabilization of the intracavity power and pump-resonance detuning [33, 40] can improve the timing jitter PSD at the low offset frequency. By increasing tapered waveguide width (increasing the effective resonant mode volume) and decreasing the cavity GVD, the jitter of the frequency DM-microcomb oscillator can be improved to sub-femtosecond timing imprecision.

**Conclusion**

In this study the fundamental noise of dispersion-managed soliton microcombs without a restoring force are examined in detail. Dispersion-managed microcombs are deterministically and reliably generated with a TE-TM dual-driven thermally stabilized approach at the single-soliton, double-soliton, and soliton crystal regimes. The RIN is determined to be -153.2 dB/Hz at 100 kHz offset for the single-soliton state, with parameters bounded by the CW pump laser. The timing jitter PSD is 0.4 as$^2$/Hz at 100 kHz offset and the corresponding integrated timing jitter is 1.7 ± 0.07 fs from 10 kHz to 1 MHz (≈ 32.3 fs from 10 kHz to Nyquist 44.5 GHz). To the best of our knowledge, we achieved femtosecond timing jitter for the first time in dispersion-managed microcombs. The demonstrated results show the dynamic noise for the double-soliton and soliton crystal in the RIN PSD is important to understand intracavity soliton dynamics, with the single-soliton state having the lowest jitter, and with slight but quantifiable variations across the different soliton states.

The primary noise source at low offset frequencies is the fluctuation in effective cavity length, which arises from intracavity power fluctuations in the microresonator. In dispersion-managed microcombs, we observe negligible center frequency shifts, which helps to prevent noise conversion processes related to center-frequency shifts. Future studies could explore how high-order dispersion in these microresonators facilitates additional noise coupling mechanisms. By implementing feedback to stabilize the pump laser's intracavity power and frequency, the timing jitter of the dispersion-managed chip-scale soliton oscillator could be reduced to sub-femtosecond levels. Furthermore, balancing factors such as detuning, nonlinearity, higher-order dispersion, and avoided mode crossing in the microresonator can further minimize timing jitter.

*Disclosures*

The authors declare that there are no financial interests, commercial affiliations, or other potential conflicts of interest that could have influenced the objectivity of this research or the writing of this paper.

*Code, Data, and Materials Availability*

Availability of code, data, and/or materials used in the research results reported in the manuscript may be provided upon reasonable request.

*Author contributions*

W.W. and H.L. conducted the experiments. W.W. and W.L. analyzed the data. W.W., H.L., and W.L. contributed to the simulations. W.L. performed the noise simulations. J.Y. designed the microresonator. T.M., D.L., J.Y., A.K.V., J.L. and Y.S.J. contributed to the design of the experiments. M.Y. and D.-L.K. performed the device nanofabrication. W.W., J.Y., P.D., J.C, and C.W.W. initiated the project. W.W. and C.W.W. wrote the manuscript. All authors discussed the results. Correspondence and requests for materials should be addressed to W.L. and C.W.W. (wzliu@g.ucla.edu; cheewei.wong@ucla.edu).

*Acknowledgments*

The authors acknowledge fruitful discussions with Prof. Shu-Wei Huang, Scott Diddams, Tara Drake, and Dohyeon Kwon on the measurements, with Prof. Guo Qing Chang and Prof. Ming Xin on noise theory, and with Prof. Ken Yang. We acknowledge financial support from the Lawrence Livermore National Laboratory (B622827), the National Science Foundation (1824568, 1810506, 1741707, 1829071), and the Office of Naval Research (N00014-16-1-2094).

*Biographies*

**Wenting Wang** received his Ph.D. degree from the Institute of Semiconductors, Chinese Academy of Sciences, Beijing, China, and Deutsches Elektronen-Synchrotron, Hamburg, Germany, in 2017. From 2017 to 2022, he was a Postdoctoral Fellow at the University of California, Los Angeles, America. He is currently a Professor at the Xiongan Institute of Innovation, Chinese Academy of Sciences, Xiong'an New Area, China. His research interests include ultralow-noise microwave/terahertz generation, micro-cavity optical frequency comb, optical signal processing, integrated microwave photonics, and high-precision time and frequency transfer.

**Wenzheng Liu** is a Ph.D. student in the Department of Electrical Engineering at the University of California, Los Angeles. He received his B.S. degree from the University of Science and Technology of China in 2018 and then received his M.S. degree from the California Institute of Technology in 2020. His research interests include micro-cavity optical frequency comb, integrated microwave photonics, quantum optics, and computer vision.

**Ninghua Zhu** received the B.S., M.S., and Ph.D. degrees in electronic engineering from the University of Electronic Science and Technology of China, Chengdu, China, in 1982, 1986, and 1990, respectively. From 1990 to 1994, he was a Postdoctoral Fellow at Sun Yat-sen University, Guangzhou, China, where he became an Associate Professor in 1992 and a Full Professor in 1994. He is currently a Professor at the Institute of Semiconductors, Chinese Academy of Sciences, Beijing, China. He has authored or co-authored more than 200 journal papers, three books, and three book chapters. He is an Associate Editor for Optics Express. His research interests include modeling and characterization of integrated optical waveguides and coplanar transmission lines and optimal design of optoelectronics devices and photonic integrated circuits.

**Chee Wei Wong** received his B.S. degree in mechanical engineering from the University of California at Berkeley, Berkeley, in 1999 and his M.S. and Ph.D. degrees from Massachusetts Institute of Technology, Cambridge, in 2001 and 2003, respectively. He was a Postdoctoral Fellow at Massachusetts Institute of Technology in 2003. He is currently a Professor at the Electrical Engineering Department at the University of California, Los Angeles. He is a Fellow of the American Physical Society. His current research interests include nonlinear and quantum optics in nanophotonics, silicon electronic-photonic circuits and photonic crystals, quantum dot interactions in nanocavities, nano-electromechanical systems, and nanofabrication.

Biographies of the other authors are not available.